\documentclass[aps,pre,preprint,endfloats]{revtex4}

\usepackage{amsfonts}
\usepackage{graphicx}

\newcommand{\sref}[1]{Sec.~\ref{#1}}

\font\bbfnt=msam10

\def\lsim{\,\hbox{\bbfnt\char'056}\,}

\date{\today}

\makeatletter
\def\@dotsep{4.5}
\makeatother
\begin{document}
\title{Quantum dynamics of a qubit coupled with structured bath}

\author{Peihao Huang}
 \email{phhuang@sjtu.edu.cn}
\author{H. Zheng}
 \affiliation{Department of Physics, Shanghai Jiao Tong
University, Shanghai 200240, P.R.China}

\keywords {dissipative system£¬damped oscillator£¬lorentz
structure£¬quantum bit}

\begin{abstract}

The dynamics of an unbiased spin-boson model with Lorentzian
spectral density is investigated theoretically in terms of the
perturbation theory based on a unitary transformation. The
non-equilibrium correlation function $P(t)$ and susceptibility
$\chi^{\prime\prime}(\omega)$ are calculated for both the
off-resonance case $\Delta\lesssim 0.5\Omega$ and the on-resonance
case $\Delta\sim \Omega$. The approach is checked by the Shiba's
relation and the sum rule. Besides, the coherent-incoherent
transition point $\alpha_c$ can be determined, which has not been
demonstrated for the structured bath by previous authors up to our
knowledge.

\end{abstract}

\maketitle

\section{Introduction}
Quantum computation has shown a lot of advantages in performing
certain tasks\cite{D. Bouwmeester,M. A. Nielsen,C. Monroe1}. As the
basis of quantum computer, quantum bit (qubit) is one of the most
attractive research topics today. Although, qubits have been
realized in microscopic systems many years ago\cite{Q. A.
Turchette,C. Monroe2,N. A. Gershenfeld}, it is difficult to
implement the desired large number of interacting qubits which would
be of practical value for computation\cite{J. E. Mooij}. Therefore,
macroscopic qubit systems, especially for the solid state circuits
systems, have aroused a lot of interests recently, not only for its
potential in realizing valuable quantum computer, but also for its
theoretical importance in understanding the boundary between
classical and quantum physics. Actually, about twenty years ago,
macroscopic quantum phenomena has already been proposed in
A.J.Leggett's pioneering work\cite{A.J.Leggett}. Through several
years' efforts, rapid progress has been made in the field of
scalable qubits, especially for the superconducting qubits including
charge\cite{charge1,charge2}, phase\cite{phase1,phase2} and flux
qubits\cite{flux1,flux2,flux3,flux4}.

Although Rabi oscillations have been observed in those qubit
systems\cite{charge1,charge2,phase1,phase2,flux3,flux4}, the
decoherence is still the biggest obstacle on the avenue toward
realizing a quantum computer today. For a quantum system, any
irreversible interaction with the environment will completely
destroy the quantum coherence. This raises the problem of exploring
the dynamics of qubit system with dissipative environment. Take flux
qubit systems for example\cite{flux3,flux4,map_T. L. Robertson}:
during measurement, the qubit is coupled with the detector, which
itself is coupled to the outside environment. Via the coupling which
extracts information from qubit system, the noise is also
transmitted to the qubit system from SQUID's noncoherent
environment. In this case, the environment affects significantly the
qubit system and has to be taken into account. In order to explore
the effect of environment on the qubit system, we can introduce an
ohmic boson bath as a description of the non-coherent environment.
And it has been proved that this kind of qubit system can be mapped
to spin-boson model and the only difference is the spectral
density\cite{map_Lin Tian,map_C. H. van der Wal,map_T. L.
Robertson}, so we only have to concentrate on the spin-boson model
with different structured baths.

In this paper, we study the dynamics of a flux qubit measured by a
dc-SQUID. After mapping the system to spin-boson model, the
corresponding spectral density is of Lorentzian form:
\begin{equation}\label{J(omega)}
J(\omega)=\frac{2\alpha\omega\Omega^4}{(\Omega^2-\omega^2)^2+(2\pi\Gamma\omega\Omega)^2}.
\end{equation}
We study this spin-boson model in terms of the perturbation
treatment based on a unitary transformation which was proposed by H.
Zheng\cite{H.Zheng-04}. This perturbation method can lead to the
analytical results for the non-equilibrium correlation function and
the susceptibility. Actually, many methods have been used to study
this kind of systems, however, a spectral density of the type
Eq.\ref{J(omega)} poses a challenge to most of these
approaches\cite{floweq&Bloch-Redfield&NIBA_F.K.Wilhem}. By
comparison with those approaches, our perturbation method
demonstrates some good features: it works well for a wide parameter
range. And coherence-decoherence transition point $\alpha_c$ can be
calculated.

This paper is organized as follows: in \sref{sec:model}, we present
the model and give an alternative view of this model. In
\sref{sec:transform}, we introduce the unitary transformation
briefly. In \sref{sec:P(t)}, we calculate the non-equilibrium
correlation function $P(t)$, compare the results with that of other
methods. In \sref{sec:chi(t)} the Green's function and
susceptibility are calculated, and Shiba's relation are validated.
In \sref{sec:alpha_c}, the coherence-decoherence transition point
$\alpha_c$ is studied in detail.

\section{Qubit-Environment Interacting Hamiltonian}\label{sec:model}

In a flux qubit system, the qubit is the two macroscopically
distinct quantum states representing clockwise and anticlockwise
rotating supercurrents. And information in the qubit is detected by
the outside circuits including a dc-SQUID\cite{flux3,flux4}. In this
case, the qubit is entangled with the detecting field, which is
itself coupled with the outside noncoherent environment. The qubit
can be characterized by a pseudospin-1/2 operator $\sigma_x$ as
usual (unbiased condition), the detecting equipment, which is
actually a LC resonant circuit\cite{flux3,flux4}, can be described
by a harmonic oscillator with a characteristic frequency $\Omega$
and the outside environment can be described by a set of harmonic
oscillators. Therefore, the qubit-meter-environment interaction
hamiltonian can be written as($\hbar=1$):
\begin{eqnarray}
H&=&-\frac{\Delta}{2}\sigma_x+\frac{P^2}{2M}+\frac{M\Omega^2}{2}(X+q\sigma_z)^2 \nonumber  \\
&+&\sum_k\left[\frac{p_k^2}{2m_k}+\frac{m_k\omega_k^2}{2}\big(x_k+\frac{c_kX}{m_k\omega_k}\big)^2\right],\label{Hamiltonian1}
\end{eqnarray}
where $\Delta$ represents the frequency of tunneling between the two
states of the qubit, $\Omega$ is the frequency of meter,
$\omega_k$'s are the frequencies of the oscillators which represent
the outside environment $(k=1,2,3,\cdots)$, $q\sigma_z$ is the
displacement of qubit caused by the interaction with meter, which
also has a displacement of $c_kX/(m_k\omega_k)$ caused by the
interaction with outside environment. Here, the coupling between
qubit and meter is assumed to be linear, and the same assumption is
applied to the coupling between meter and outside
environment\cite{floweq_Silvia Kleff}.

According to the second quantization process, the Hamiltonian $H$
can be written as:
\begin{eqnarray}
\widetilde{H} & = &-\frac{\Delta}{2}\sigma_x+{\Omega}B^{\dagger}B+
\sum_k\widetilde{\omega}_k\tilde{b}^{\dagger}\tilde{b} \nonumber \\
&+&(B^{\dagger}+B)\bigg[g\sigma_z
 +\sum_k\kappa_k(\widetilde{b}^{\dagger}_k+\widetilde{b}_k)\bigg]
 +(B^{\dagger}+B)^2\sum_k\frac{\kappa_k^2}{\widetilde{\omega}_k},\label{Hamiltonian2}
\end{eqnarray}
where $B$ (or $B^\dag$) and $\tilde{b}_k$ (or $\tilde{b}_k^\dag$)
are the annihilation (or creation) operators of harmonic oscillators
with frequencies $\Omega$ and $\omega_k$'s, respectively. $g$ and
$\kappa_k$ are the coupling constants.The coupling to the
environment is fully defined by the spectral density,which is
usually taken to be of ohmic form to model the dissipative
environment: $
\tilde{J}(\omega)\equiv\sum_k\kappa_k^2{\delta}(\omega-\widetilde{\omega}_k)=\Gamma\omega\theta(\omega_c-\omega)
$.

The Hamiltonian (\ref{Hamiltonian2}) can be mapped to spin boson
model \cite{map_Lin Tian,J.Chem.Phys-85}:
\begin{equation}
H=-\frac{\Delta}{2}\sigma_x+\sum_k{\omega_k}b^{\dagger}_kb_k+\frac{1}{2}\sigma_z\sum_kg_k(b_k^{\dagger}+b_k),\label{Hamiltonian3}
\end{equation}
where the spin dynamics depends only on the Lorentzian structured
spectral density which is approximately given by:
\begin{equation}
J(\omega)\equiv\sum_kg_k^2{\delta}(\omega-{\omega}_k)=\frac{2\alpha\omega\Omega^4\theta(\omega_c-\omega)}{(\Omega^2-\omega^2)^2+(2\pi\Gamma\omega\Omega)^2},
\end{equation}
where $\alpha=8{\Gamma}g^2/{\Omega^2}$. Since the cutting
frequency $\omega_c$ is always much larger than $\Omega$, the
spectral density $J(\omega)$ can be reasonably taken as
Eq.(\ref{J(omega)}). When the characteristic frequency $\Omega$ is
higher than others, say $\Omega > 2\Delta$, the Lorentzian
structured spectral density is nearly the same as the Ohmic
spectral density which has been extensively studied. This case
will be called as off-resonance. For the on-resonance case $\Omega
\sim \Delta$, the physical properties of the coupling system with
Lorentzian structured spectral density  may be quite different
from those of the Ohmic bath.

\section{Unitary Transformation}\label{sec:transform}
Now we apply an unitary transformation to the Hamiltonian
(\ref{Hamiltonian3})\cite{H.Zheng-04}:$
H^{\prime}=\exp(S)H\exp(-S)$, where
\begin{equation}
S\equiv\sum_k\frac{g_k}{2\omega_k}\xi_k(b^{\dag}_k-b_k)\sigma_z.
\end{equation}
If $\xi_k=1$, the generator S reduces to the usual polaron
transformation. After the unitary transformation, the Hamiltonian
can be decomposed into three parts:
\begin{eqnarray}
H^{\prime}=H_0^{\prime}+H_1^{\prime}+H_2^{\prime},
\end{eqnarray}
where
\begin{eqnarray}
    H^{\prime}_0&=&-\frac{\sigma_x}{2}\eta\Delta+\sum\limits_k{\omega_k}b_k^{\dagger}b_k-\sum\limits_k\frac{g_k^2}{4\omega_k}\xi_k(2-\xi_k)\label{H_0'},\\
    H_1^{\prime}&=&\frac{\sigma_z}{2}\sum\limits_kg_k(1-\xi_k)(b_k^{\dagger}+b_k)\nonumber\\
    &&-\frac{i\sigma_y}{2}\eta\Delta\sum\limits_k\frac{g_k}{\omega_k}\xi_k(b_k^\dag-b_k),\label{formerH_1'}\\
    H_2^{\prime}&=
    &-\frac{\sigma_x}{2}\Delta\left\{\cosh\left[\sum\limits_k\frac{g_k}{\omega_k}\xi_k(b_k^\dag-b_k)\right]-\eta\right\}\nonumber\\
    &
    &-\frac{i\sigma_y}{2}\Delta\bigg\{\sinh\left[\sum\limits_k\frac{g_k}{\omega_k}\xi_k(b_k^\dag-b_k)\right]
    -\eta\sum\limits_k\frac{g_k}{\omega_k}\xi_k(b_k^\dag-b_k)\bigg\}\label{H_2'}.
\end{eqnarray}

Obviously, $H_0^{\prime}$ can be solved exactly since the spin and
bosons are decoupled in $H_0^{\prime}$. The eigenstates of
$H_0^{\prime}$ can be expressed as a direct product:
$|s\rangle|\{n_k\}\rangle$, where $|s\rangle$ is the eigenstates of
$\sigma_x$, which can be $|s_1\rangle$ or $|s_2\rangle$, and
$|\{n_k\}\rangle$ is the eigenstates of phonons, which means that
there are $n_k$ phonons for mode $k$. Therefore, the ground state of
$H_0^{\prime}$ is given by :
$|g_0\rangle=|\,s_1{\rangle}|\{0_k\}\rangle,$ where
$|\,s_1{\rangle}$ is the lower eigenstate of spin and
$|\{0_k\}\rangle$ stands for the vacuum state for phonons. Also the
lowest excited states can be denoted as
$|\,s_2{\rangle}|\{0_k\}\rangle$ and
$|\,s_1{\rangle}|\{1_k\}\rangle$ where $|\{1_k\}\rangle$ is the
number state with $n_k=1$ but $n_{k^{\prime}}=0$ for all
$k^{\prime}\neq k$.

Since $H_1^{\prime}$ and $H_2^{\prime}$ will be treated as
perturbation, they should be as small as possible.  In order to
minimize $H_1^{\prime}$ and $H_2^{\prime}$, we let
$H_1^{\prime}|g_0\rangle=0$ and
${\langle}g_0|H_2^{\prime}|g_0\rangle=0$. Then, the parameters
$\eta$ and $\xi_k$'s are determined as,
\begin{equation}\label{eta}
\eta=\exp\left[-\sum\limits_k\frac{g_k^2}{2\omega_k^2}\xi_k^2\right],
\end{equation}
\begin{equation}\label{xi_k}
\xi_k=\frac{\omega_k}{\omega_k+\eta\Delta}.
\end{equation}
Note that $0\leq\xi_k\leq1$ measures the intensity of the spin-boson
coupling: $\xi_k\sim1$ if the boson frequency ¦Øk is larger than the
renormalized tunnelling $\eta\Delta$; but  $\xi_k\ll1$ for
$\omega_k\ll\eta\Delta$. Since the transformation generated by S is
a displacement, physically, one can see that high-frequency bosons
($\omega_k>\eta\Delta$) follow the tunnelling particle adiabatically
because the displacement is $g_k\xi_k/\omega_k\sim{g_k/\omega_k}$.
However, bosons of low-frequency modes $\omega_k<\eta\Delta$ in
general are not always in equilibrium with the tunnelling particle,
and hence the particle moves in a retarded potential arising from
the low-frequency modes. When the non-adiabatic effect dominates,
$\omega_k\ll\eta\Delta$, the displacement $\xi_k\ll1$.

The elements of Hamiltonian matrix of ground and lowest excited
states can be written as:
\begin{equation}\label{H'_Matrix}
\begin{array}{c|cccc}
                               &  |g_0\rangle            & |\psi_0\rangle & |\psi_k\rangle & |\psi_{k^{\prime}}\rangle  \\
\hline
{\langle}g_0| & -\frac{\eta\Delta}{2} &               0            &                     0                        &                0\\
{\langle}\psi_0| &             0           &   \frac{\eta\Delta}{2}   &V_k & V_{k^{\prime}}\\
{\langle}\psi_k| &             0           &V_k                         &-\frac{\eta\Delta}{2}+\omega_k&0\\
{\langle}\psi_{k^{\prime}}| &             0 &V_{k^{\prime}} & 0
&-\frac{\eta\Delta}{2}+\omega_{k^{\prime}}
\end{array}\nonumber
\end{equation}
where $H^{\prime}_2$ has been dropped, $|\psi_0\rangle$ and
$|\psi_k\rangle$ (k=1,2,$\cdots$) are the lowest exited states which
represent $|\,s_2\rangle|\{0\}\rangle$ and
$|\,s_1\rangle|\{1_k\}\rangle$, respectively.
$V_k=\eta{\Delta}g_k\xi_k/\omega_k={g_k\eta\Delta}/{(\omega_k+\eta\Delta)}$.

Since $|g_0\rangle$ has already been diagonalized, we can now
diagonalize the lowest excited states of $H^{\prime}$ as follows:
\begin{eqnarray}
  H^{\prime}&=&-\frac{\eta\Delta}{2}|g_0\rangle{\langle}g_0|+\sum\limits_{E}E|E\rangle{\langle}E|\label{transformation}\\
  &&+\texttt{ terms with higher excited states.}\label{H'_Phi}\nonumber
\end{eqnarray}
The transformation is given by:
\begin{eqnarray}
|E\rangle\,&=&x(E)|\psi_0\rangle+\sum\limits_{k}y_k(E)|\psi_k\rangle,\label{E_psi_final}\\
|\psi_0\rangle&=&|\,s_2\rangle|\{0\}\rangle=\sum\limits_Ex(E)|E{\rangle},\\
|\psi_k\rangle&=&|\,s_1\rangle|\{1_k\}\rangle=\sum\limits_Ey_k(E)|E\rangle,
\end{eqnarray}
where
\begin{eqnarray}
x(E)\;&=&\bigg[1+\sum\limits_{k\neq0}\frac{V_k^2}{(E+\frac{\eta\Delta}{2}-\omega_k)^2}\bigg]^{-\frac{1}{2}},\quad\label{x(E)}\\
y_k(E)&=&\frac{V_k}{E+\frac{\eta\Delta}{2}-\omega_k}x(E),\label{y_k(E)}
\end{eqnarray}
where E's are the diagonalized excitation energy and they are the
solutions of eigenvalue equation:
\begin{equation}\label{E_equation}
E-\frac{\eta\Delta}{2}-\sum\frac{V_k^2}{E+\frac{\eta\Delta}{2}-\omega_k}=0.
\end{equation}

\section{The Non-Equilibrium Correlation Function}\label{sec:P(t)}

The non-equilibrium correlation function $P(t)$ is defined as
$P(t)={\langle}\phi(t)|\sigma_z|\phi(t)\rangle$, and
$|\phi(t)\rangle$ is the wave function in the Schr\"{o}dinger
picture:
\begin{equation}\label{wave function}
|\phi(t)\rangle=e^{-iHt}|\phi(0)\rangle.
\end{equation}
we choose the initial state as $|+\rangle|b,+\rangle$ where
$|+\rangle$ is the eigenstate of $\sigma_z=+1$ and $|b,+\rangle$ is
the state of bosons adjusted to the state of $\sigma_z=+1$. Because
of the unitary transformation, the non-equilibrium correlation
function $P(t)$ can be written as:
\begin{equation}\label{P(t)_transform}
P(t)=\langle\{0\}|\langle+|e^{iH{^{\prime}}t}\sigma_ze^{-iH{^{\prime}}t}|+\rangle|\{0\}\rangle.
\end{equation}
By using Eqs.\ref{transformation}-\ref{E_equation}, we can get:
\begin{eqnarray}
P(t)&=&\frac{1}{2}\sum_Ex^2(E)\left(e^{-i(E+\frac{\eta\Delta}{2})t}+e^{i(E+\frac{\eta\Delta}{2})t}\right)\nonumber\\
&=&\frac{1}{4{\pi}i}\bigg\{\oint\frac{dE{^{\prime}}e^{-iE{^{\prime}}t}}{E^{\prime}-\eta\Delta-R(\omega)+i\gamma(\omega)}\nonumber\\
&&+\oint\frac{dE{^{\prime}}e^{iE{^{\prime}}t}}{E^{\prime}-\eta\Delta-R(\omega)-i\gamma(\omega)}\bigg\},\label{P(t)_int}
\end{eqnarray}
where a change of the variable $E^{\prime}=E+\eta\Delta/2$ is made
and residue theory has been used. The contour of the first integrand
in Eq.\ref{P(t)_int} is composed of a straight line which is
infinitesimally close to the real axis from above and a semicircle
above the real axis with infinite radius. And the contour of the
second one is composed of a straight line which is infinitesimally
close to the real axis from below and a semicircle under the real
axis with infinite radius. $R(E^{\prime})$ and
$\pm\gamma(E^{\prime})$ in Eq.\ref{P(t)_int} are the real and
imaginary parts of $\sum_kV_k^2/(E^{\prime}{\pm}i0^+-\omega_k)$ and
they can be written as:
\begin{eqnarray}
R(\omega)=(\eta\Delta)^2\int_0^{\infty}\,d\omega^{\prime}\frac{J(\omega^{\prime})}{(\omega^{\prime}+\eta\Delta)^2(\omega-\omega^{\prime})},\label{R(omega)}\\
\gamma(\omega)=\pi(\eta\Delta)^2J(\omega)/(\omega+\eta\Delta)^2.\quad\quad\quad\quad\;\;\label{gamma(omega)}
\end{eqnarray}

The integral in Eq.(\ref{P(t)_int}) can proceed by calculating the
residue of integrand and the result is:
\begin{equation}\label{P(t)_coswt}
    P(t)=e^{-{\gamma}t}\cos(\omega_0t),
\end{equation}
where $\omega_0$ is the solution of equation:
\begin{equation}\label{omega_0}
\omega-\eta\Delta-R(\omega)=0,
\end{equation}
and $\gamma=\gamma(\eta\Delta)={\pi}J(\eta\Delta)/4$ where we have
applied the second order approximation\cite{Quantum optics}. Once
the parameters $\alpha$, $\Gamma$ and $\Omega$ of the system are
given, the renormalized frequency of the tunnelling between two
states can be derived from Eq.(\ref{omega_0}).

\begin{figure}
\centering
  \includegraphics[scale=2.2]{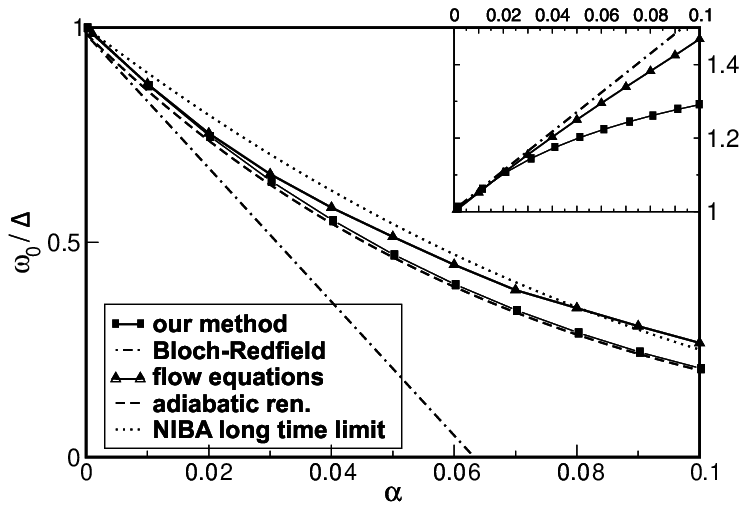}%
  \caption{The renormalized tunneling frequency $\omega_0$
verses the coupling strength $\alpha$. Main plot:
$\Delta/\Omega$=0.1, $\Gamma$=0.02; inset: $\Delta/\Omega$=1.1,
$\Gamma$=0.06. The results of the other methods are from
F.K.Wilhem's
article\cite{floweq&Bloch-Redfield&NIBA_F.K.Wilhem}.}\label{Fig.omega_0-alpha}\clearpage
\end{figure}

For checking our approach, we calculate the renormalized tunnelling
frequency $\omega_0$ for both the off- and on-resonance cases and
make comparison with results of the other methods. The main plot in
Fig.\ref{Fig.omega_0-alpha} describes the off-resonance case. We can
see that the tunnelling frequency $\omega_0$ decreases as the
coupling strength $\alpha$ increases. This is the similar behavior
as that of Ohmic case because the Lorentzian structured spectral
density becomes the ordinary Ohmic one for $\Delta/\Omega\ll 1$. But
it is quite different in the on-resonant case with
$\Delta\sim\Omega$ (the inset of Fig.2). We can see that as the
coupling strength $\alpha$ increases, the tunnelling frequency
$\omega_0$ increases too. It is said that the coupling enhances the
tunnelling frequency\cite{floweq&Bloch-Redfield&NIBA_F.K.Wilhem}.

\begin{figure}
\centering
  \includegraphics[scale=1.6]{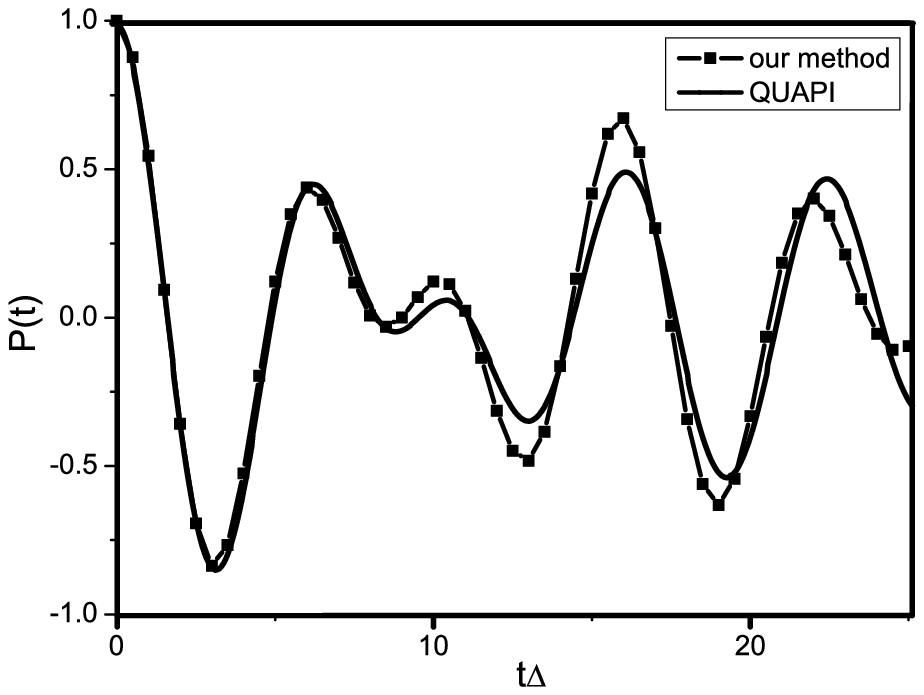}
  \caption{The dynamics of non-equilibrium correlation
  function $P(t)$. The parameters are: $\alpha=0.004$, $\Delta=\Omega$, $\Gamma=0.014$.
  The result of quasiadiabatic propagator path-integral (QUAPI) method are from M.Thorwart's
  article.\cite{QUAPI_M. Thorwart}}
  \label{Fig.compare}\clearpage
\end{figure}

Another way for calculating $P(t)$, which may be more precise, is
to do the investigation in Eq.(22) directly with the help of
Kramers-Kronig relation:
\begin{eqnarray}
P(t)&=&\frac{1}{\pi}\int_0^\infty\frac{d\omega\,\gamma(\omega)\cos({\omega}t)}{[\omega-\eta\Delta-R(\omega)]^2+\gamma^2(\omega)}.\label{int_P(t)2}
\end{eqnarray}
Fig.\ref{Fig.compare} shows our calculation, which is compared
with the result of quasiadiabatic propagator path-integral (QUAPI)
method. Note that it is in the on-resonance case and one can see
that $P(t)$ shows a double-frequency oscillation.
\begin{figure}
\centering
  \includegraphics[scale=1.4]{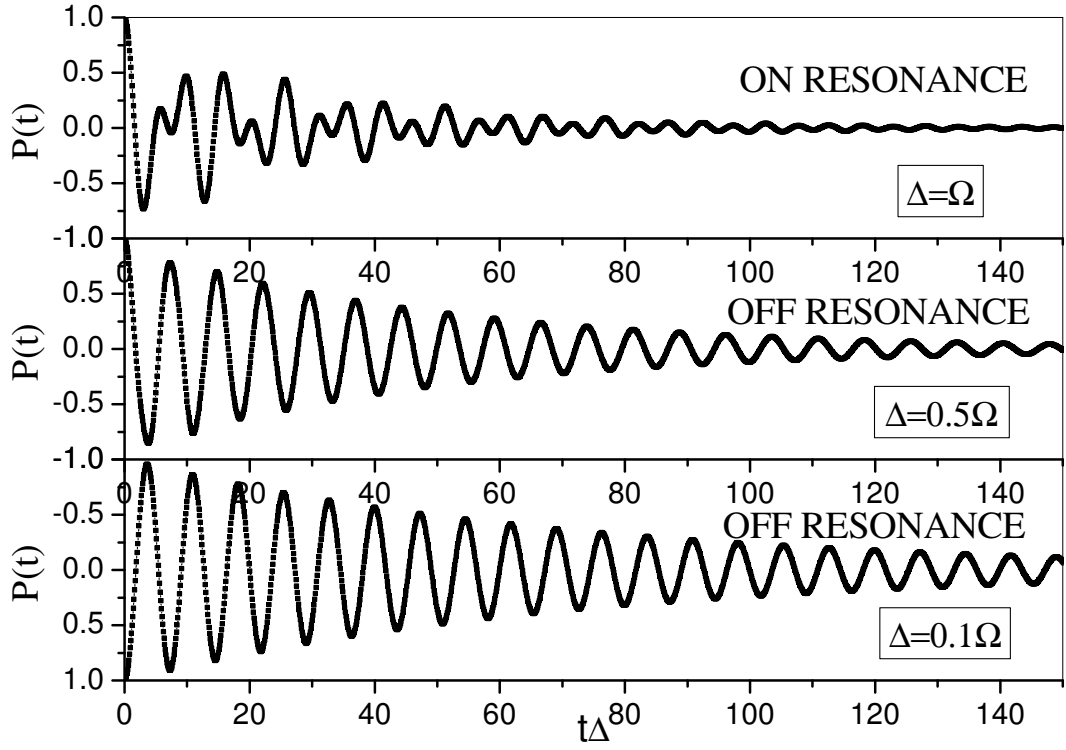}
  \caption{The dynamics of non-equilibrium correlation
  function $P(t)$. The parameters are: $\alpha=0.01$, $\Gamma=0.02$.}
  \label{Fig.Ptsa}\clearpage
\end{figure}
\begin{figure}
\centering
  \includegraphics[scale=1.4]{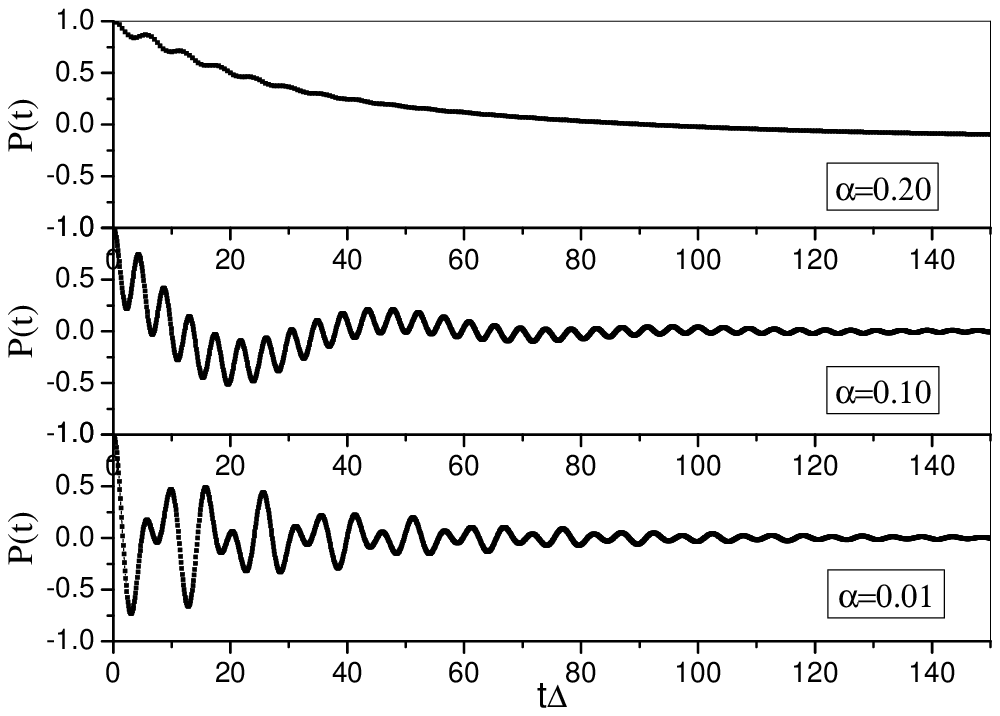}
  \caption{The dynamics of non-equilibrium correlation
  function $P(t)$ for the on-resonance case. The parameters are: $\Delta=\Omega$, $\Gamma=0.02$.}
  \label{Fig.PtsD}\clearpage
\end{figure}

Fig.\ref{Fig.Ptsa} shows the behavior of $P(t)$ for different ratio
$\Delta/\Omega$. It is a single-frequency oscillation in
off-resonance case ($\Delta/\Omega\lesssim 0.5$). However, in the
on-resonance region ($\Delta\sim\Omega$), $P(t)$ has two
characteristic frequencies. Fig.\ref{Fig.PtsD} shows the dynamics of
non-equilibrium correlation function $P(t)$ for the on-resonance
case for different coupling. Its behavior changes significantly as
$\alpha$ increases and becomes an over-damping curve as $\alpha$
becomes large enough.

The Fourier transformation of $P(t)$ is given by:
\begin{equation}
P(\omega)=\frac{1}{2\pi}\frac{\gamma(\omega)}{[\omega-\eta\Delta-R(\omega)]^2+\gamma^2(\omega)},\label{P(omega)}\\
\end{equation}
\begin{figure}
\begin{center}
\centering
  \includegraphics[scale=1.4]{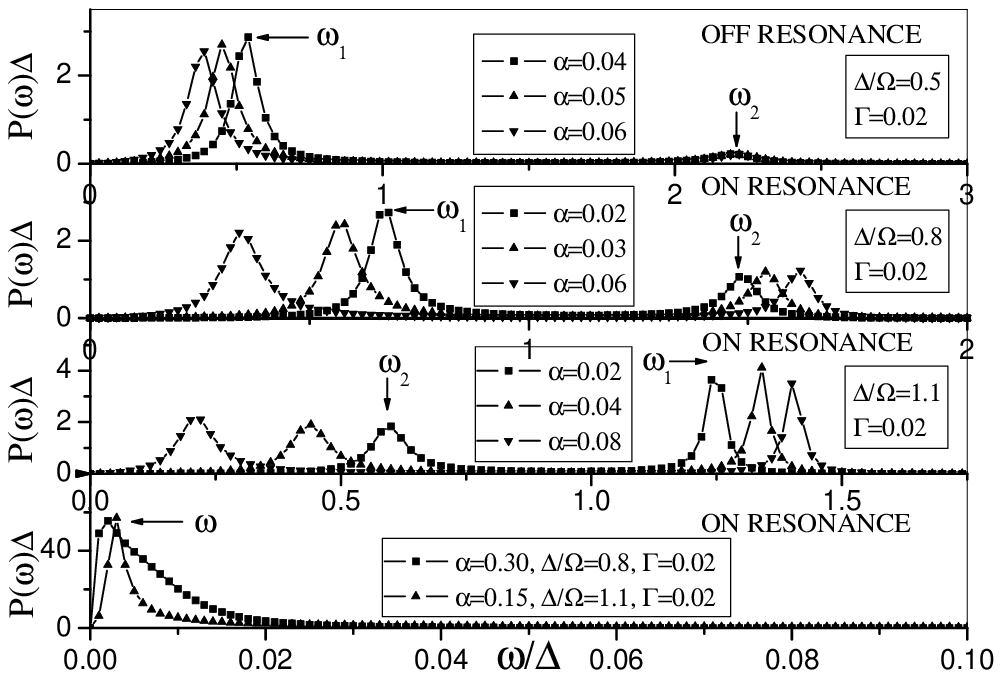}
  \caption{$P(\omega)$ verses $\omega$ for different parameters. $P(\omega)$ always has two inherent frequencies
  $\omega_1$ and $\omega_2$ corresponding to $\Delta$ and $\Omega$. First figure shows $\omega_1$ is the dominant
  frequency for the off-resonance cases ($\Delta/\Omega\lsim 0.5$). Second and third
  figure show that $\omega_1$ and $\omega_2$ are both important for
  the on-resonance cases ($\Delta\approx\Omega$) when $\alpha$ is small. In addition, $\omega_1$
  decreases and $\omega_2$ increases as $\alpha$ increases when
  $\Delta<\Omega$. $\omega_1$ increases and $\omega_2$ decreases as
  $\alpha$ increases when $\Delta>\Omega$. Last figure shows
  that no matter $\Delta$ is smaller or larger than $\Omega$, the smaller
  one of $\omega_1$ and $\omega_2$ is the dominant frequency and close to zero when coupling is strong.
  }
  \label{Fig.PW}\clearpage
\end{center}
\end{figure}
and is shown for off- and on-resonance case in Fig.\ref{Fig.PW}.
$P(t)$ has two characteristic frequencies $\omega_1$ and $\omega_2$,
where $\omega_1$ corresponds to the tunnelling $\Delta$ and
$\omega_2$ to the detecting frequency $\Omega$. Since qubit is
entangled with the detecting system, it is reasonable that the
dynamic of qubit shows some feature of the detector. The results
also show that: For off-resonance case $\Delta\lesssim 0.5\Omega$,
we have $\omega_1<\omega_2$ and $\omega_1$ decreases as the coupling
$\alpha$ increases. When $\Delta\sim\Omega$, we have
$\omega_1>\omega_2$ and $\omega_1$ increases as the coupling
$\alpha$ increases. All these results are consistent with previous
authors.

\section{The Susceptibility and Shiba's relation}\label{sec:chi(t)}

\begin{figure}
\centering
  \includegraphics[scale=1.4]{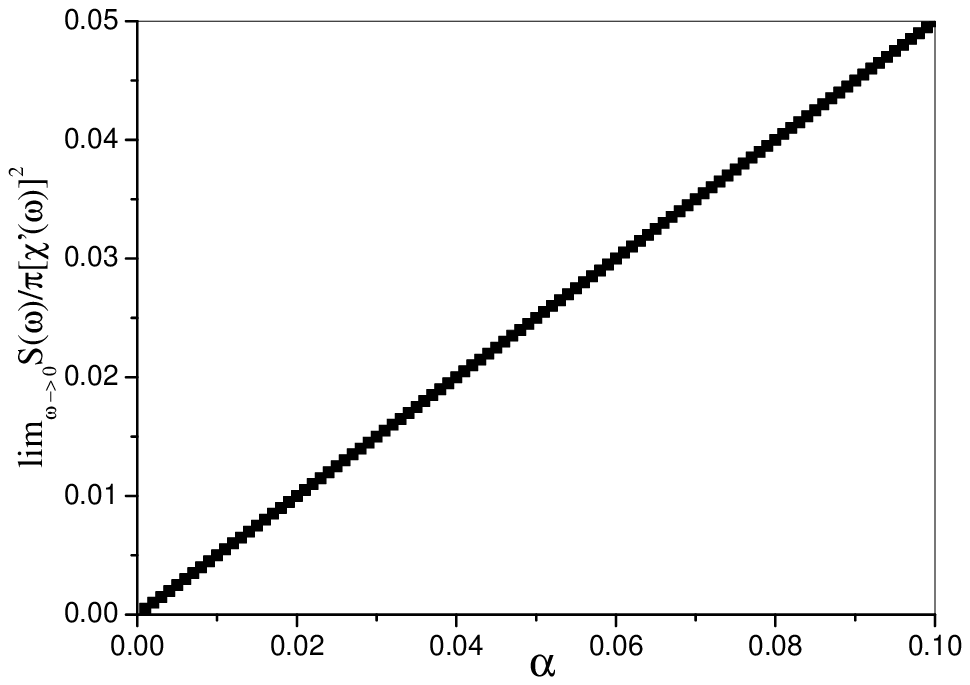}
  \caption{The validation of Shiba's relation.}
  \label{Fig.Shiba}\clearpage
\end{figure}

The retarded Green's function is defined as:
\begin{equation}\label{G(t)}
G(t)=-i\theta(t)\langle[\sigma_z(t),\sigma_z]\rangle_\beta,
\end{equation}
where $\langle\,\cdots\rangle_\beta$ means the average with
thermodynamic probability $\exp(-{\beta}H^{\prime})$ and $[A,B]$ is
the commutator $AB-BA$. The Fourier transformation of $G(t)$ is
denoted as $G(\omega)$ which satisfies an infinite chain of equation
of motion. We make the cutoff approximation for the equation chain
at the second order of $g_k$ and the solution for $T=0$
is\cite{H.Zheng-04}
\begin{eqnarray}
G(\omega)&=&\frac{1}{\omega-\eta\Delta-\sum_k{V_k^2}/(\omega-\omega_k)}\nonumber\\
         &-&\frac{1}{\omega+\eta\Delta-\sum_k{V_k^2}/(\omega+\omega_k)}.\label{G(omega)}
\end{eqnarray}
The susceptibility $\chi(\omega)=-G(\omega)$, and its imaginary part
is:
\begin{eqnarray}
\chi^{\prime\prime}(\omega)&=&\frac{\gamma(\omega)\theta(\omega)}{[\omega-\eta\Delta-R(\omega)]^2+\gamma^2(\omega)}\nonumber\\
         &-&\frac{\gamma(-\omega)\theta(-\omega)}{[\omega+\eta\Delta+R(-\omega)]^2+\gamma^2(-\omega)}.\label{chi'}
\end{eqnarray}
Define function $S(\omega)$ as:
$S(\omega)={\chi}^{\prime\prime}(\omega)/\omega$ with its limit at
$\omega\to 0$:
\begin{equation}
\lim_{\omega\to0}S(\omega)=\frac{2\pi\alpha}{[\eta\Delta+R(0)]^2}.
\end{equation}
Besides, the static susceptibility can be obtained from the
imaginary part by the Kramers-Kronig relation:
\begin{eqnarray}
\chi^{\prime}(\omega=0)&=&\frac{1}{\pi}\int_{-\infty}^{\infty}\frac{\chi^{\prime\prime}(\omega)}{\omega}d\,\omega.
\end{eqnarray}
Fig.\ref{Fig.Shiba} shows that the Shiba's
relation\cite{Shiba_1,Shiba_2,Shiba_3,Shiba_4}
\begin{equation}
\lim_{\omega\to0}S(\omega)=\frac{\pi}{2}\alpha[\chi^{\prime}(\omega=0)]^2
\end{equation}
is exactly satisfied. This fact is also a check for our approach.

\section{The Coherent-Incoherent Transition}\label{sec:alpha_c}

\begin{figure}
\centering
  \includegraphics[scale=1.6]{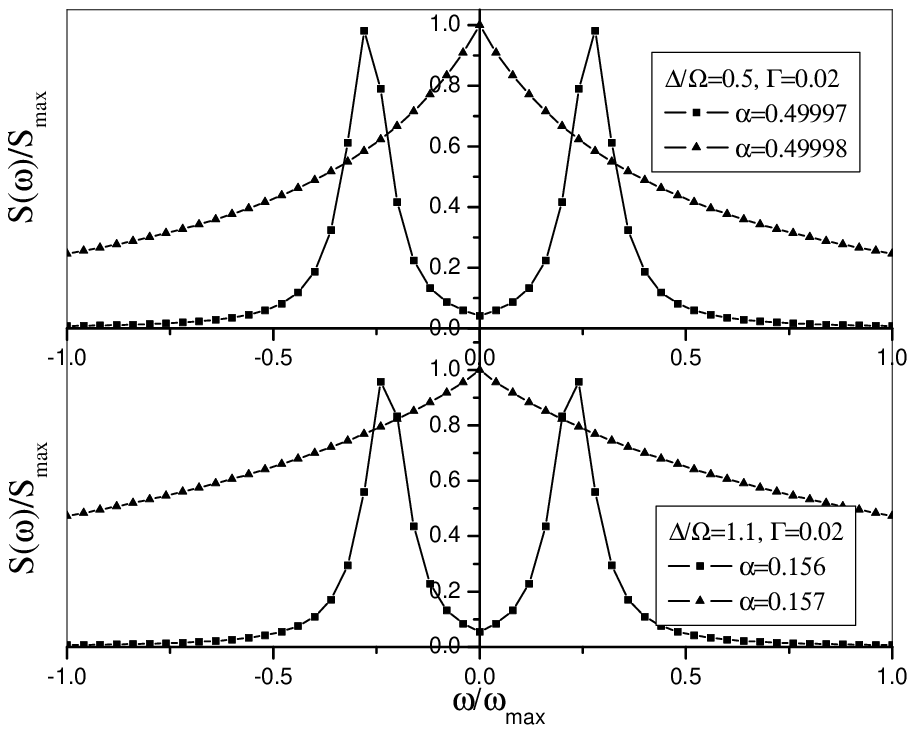}
  \caption{$S(\omega)$ verses $\omega$ for different cases.}
  \label{Fig.S(w)}\clearpage
\end{figure}

For the coherent oscillation, $S(\omega)$ has a double peak
structure symmetrical with respect to $\omega=0$. However, as soon
as the system becomes incoherent, $S(\omega)$ would have only a
quasi-elastic peak at around $\omega=0$ . Therefore, the
coherent-incoherent transition point $\alpha_c$ can be determined by
investigating the behavior of $S(\omega)$. The off-resonance case
where $\Delta/\Omega\lesssim 0.5$ and the on-resonance case where
$\Delta\sim\Omega$ will be treated separately. From
Fig.\ref{Fig.S(w)}, we can see that the coherent-incoherent
transition point is at $\alpha_c=0.49996$ when $\Delta/\Omega=0.5$,
$\Gamma=0.02$, but it is at $\alpha_c=0.156$ when
$\Delta/\Omega=1.1$, $\Gamma=0.02$. It has been validated that the
sum rule, $P(t=0)=1$ by the integration of Eq.(27), is always
satisfied when $\alpha<\alpha_c$.

\subsection{off-resonance case: $\Delta/\Omega\lesssim 0.5$}

\begin{figure}
\centering
  \includegraphics[scale=1.6]{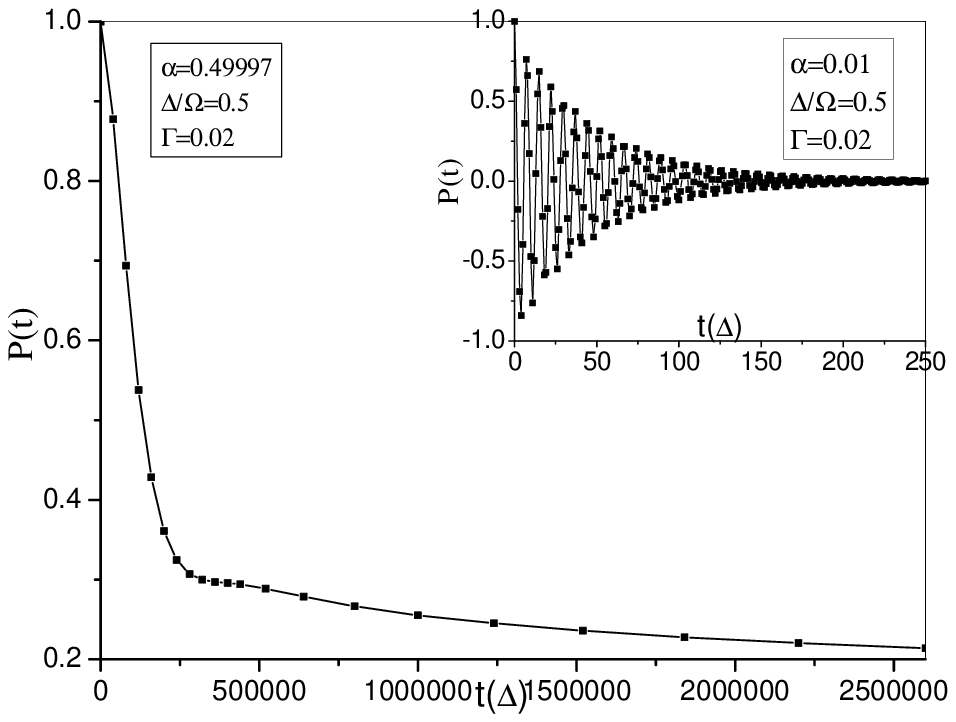}
  \caption{The dynamics of non-equilibrium correlation
  function $P(t)$ for off-resonance case. main plot: $P(t)$ is very like a over damping curve. The parameter is $\alpha=\alpha_c=0.49997$.
  Inset: Another $P(t)$ for comparison, where one inherent frequency is dominating the dynamics of non-equilibrium correlation function.
  The parameter is: $\alpha=0.01$. Other parameters are: $\Delta/\Omega=0.5$, $\Gamma=0.02$.}
  \label{Fig.Pt1}\clearpage
\end{figure}

In the off-resonance case where $\Delta/\Omega=0.5$ and
$\Gamma=0.02$, according to Fig.\ref{Fig.S(w)}, we have
$\alpha_c=0.49997$. When $\alpha$ goes to $\alpha_c$, the inherent
frequency $\omega_1$, which corresponds to $\Delta$, becomes the
dominant frequency of $P(t)$ and goes to 0. In addition, from
Eq.\ref{int_P(t)2} we have: $P(0)\approx0.999999$ for
$\alpha=\alpha_c$, where sum rule is still satisfied. Further more,
according to Fig.\ref{Fig.Pt1}, we can also find the dynamics of
non-equilibrium correlation function $P(t)$ is very like a over
damping curve at the transition point $\alpha=\alpha_c$. All these
results show that coherent-incoherent transition occurs when
$\alpha=0.49996$ for $\Delta/\Omega=0.5$ and $\Gamma=0.02$.

\subsection{on-resonance case: $\Delta\sim\Omega$}

\begin{figure}
\centering
  \includegraphics[scale=1.6]{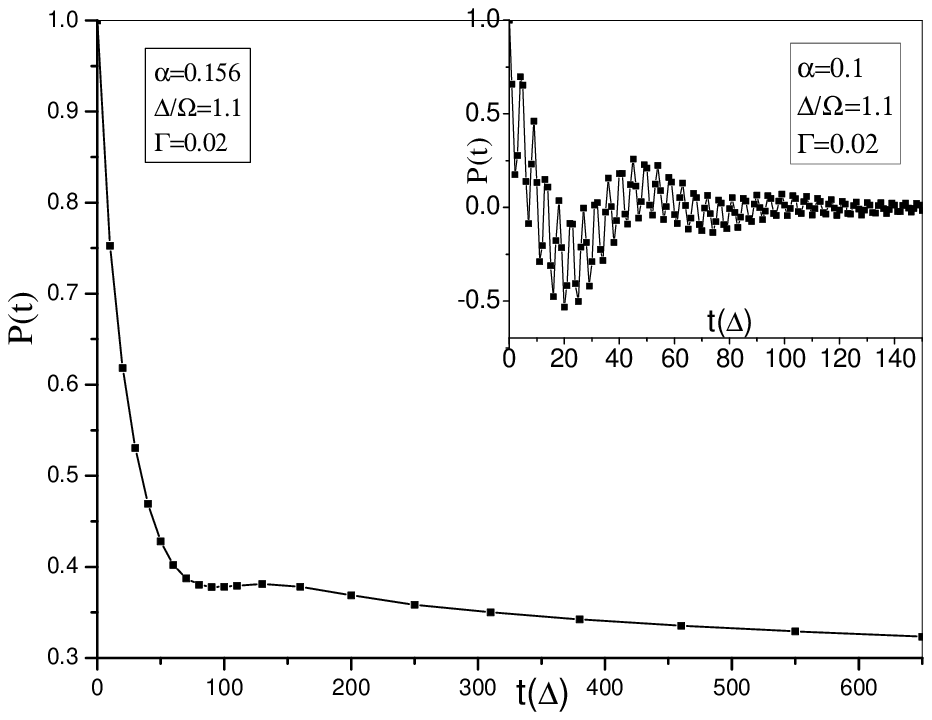}
  \caption{The dynamics of non-equilibrium correlation
  function $P(t)$ for on-resonance case. main plot: $P(t)$ is very like a over damping curve. The parameter is $\alpha=\alpha_c=0.156$.
  Inset: Another $P(t)$ for comparison, where two inherent frequency are dominating the dynamics of non-equilibrium correlation function.
  The parameter is: $\alpha=0.1$. Other parameters are: $\Delta/\Omega=1.1$, $\Gamma=0.02$.}
  \label{Fig.Pt2}\clearpage
\end{figure}

In the on-resonance case with $\Delta/\Omega=1.1$ and $\Gamma=0.02$,
Fig.\ref{Fig.S(w)} shows that $\alpha_c=0.156$. When $\alpha$
increases to $\alpha_c$, the inherent frequency $\omega_2$, which
corresponds to $\Omega$, becomes the dominant frequency of the
non-equilibrium correlation function, and goes to 0. In addition,
the integration Eq.(\ref{int_P(t)2}) leads to $P(0)\approx0.999999$
for $\alpha=\alpha_c$ where sum rule is still satisfied.
Furthermore, Fig.\ref{Fig.Pt2} shows that the dynamics of
non-equilibrium correlation function $P(t)$ looks like a over
damping curve. These results indicate that coherent-incoherent
transition is at $\alpha=0.156$ for $\Delta/\Omega=1.1$ and
$\Gamma=0.02$.

\subsection{coherent-incoherent transition point $\alpha_c$}

\begin{figure}
\centering
  \includegraphics[scale=1.6]{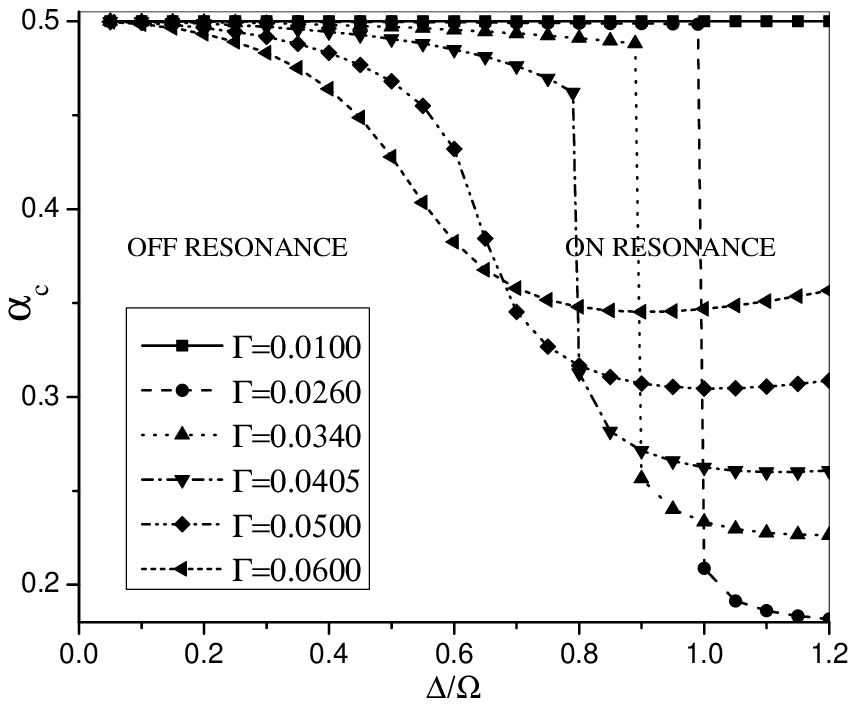}
  \caption{The coherent-incoherent transition point $\alpha_c$ verses $\Delta/\Omega$ for different $\Gamma$.}
  \label{Fig.alpha_c}\clearpage
\end{figure}

By calculating $S(\omega)$, the coherent-incoherent transition
point $\alpha_c$ can be determined as shown in
Fig.\ref{Fig.alpha_c}. $\alpha_c=1/2$ at the scaling limit
$\Delta/\Omega\ll 1$, which is the same as was predicted by
previous authors in case of ohmic bath. As the system deviates
from the scaling limit, the coherent-incoherent transition point
$\alpha_c$ is always less than 1/2, which is different from the
ohmic case where $\alpha_c$ is always larger than 1/2 for finite
$\Delta$.\cite{H.Zheng-04}

\begin{figure}
\centering
  \includegraphics[scale=1.6]{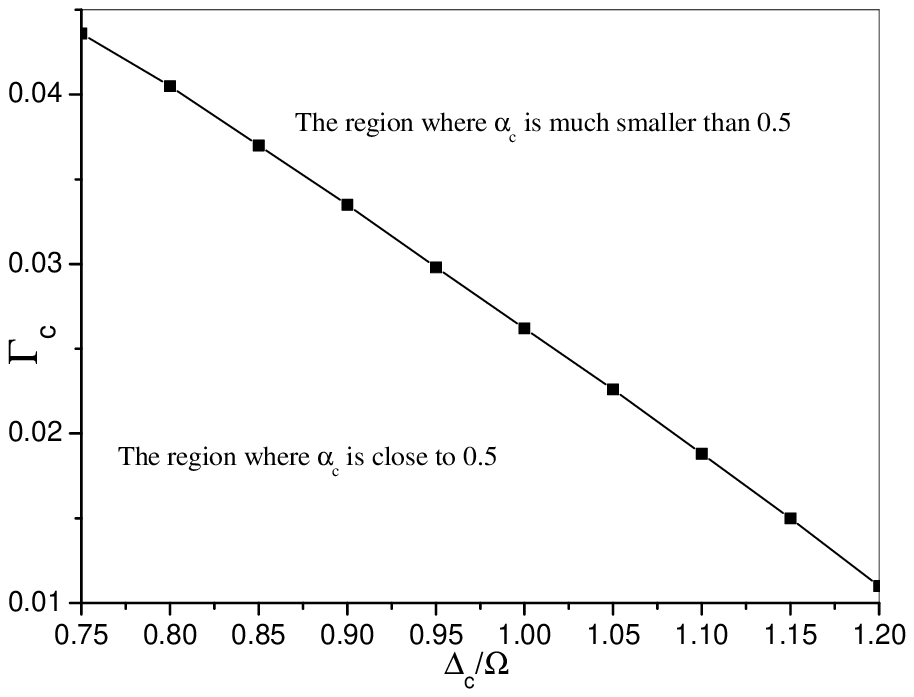}
  \caption{$\Gamma_c$ verses $\Delta/\Omega$ for on-resonance case.}
  \label{Fig.Gc}\clearpage
\end{figure}

In the off-resonance case, $\alpha_c$ decreases smoothly as
$\Delta/\Omega$ increases, and the descending becomes faster with
increasing $\Gamma$. Whereas, in the on-resonance case, there is a
suddenly drop of $\alpha_c$ to a much smaller value at some
particular values of $\Delta_c/\Omega$ and $\Gamma_c$. Our
explanation is that as $\Delta$ approaches $\Omega$, the
qubit-oscillator system becomes more and more on-resonant and the
outside incoherent factor will become more and more easier to
transfer into the qubit system under the help of oscillator
$\Omega$. So, when $\Delta/\Omega$ and $\Gamma$ reaches a particular
on-resonant point $\Delta_c/\Omega$ and $\Gamma_c$, the qubit system
will show a sudden fall of coherence. Fig.\ref{Fig.Gc} shows that
$\Gamma_c$ decreases linearly as $\Delta_c/\Omega$ increases.
Therefore, in order to maintain coherent dynamics (large
$\alpha_c$), we have to limit $\Gamma$ and $\Delta/\Omega$ in a
particular range in the on-resonance case.

\section{Conclusion}

The dynamics of an unbiased spin-boson model with Lorentzian
spectral density is investigated through a perturbation method based
on a unitary transformation. An alternative view of the system is a
two state system coupled to a single harmonic oscillator with
frequency $\Omega$, the latter being weakly coupled to an Ohmic
bath. By comparing with others, our approach shows some advantages:
it works well for both the off-resonance case $\Delta\lesssim
0.5\Omega$ and the on-resonance case $\Delta\sim \Omega$, and the
coupling constant $\alpha$ may be as large as the
coherence-incoherence transition point $\alpha_c$. We calculate the
non-equilibrium correlation function $P(t)$ and the susceptibility
$\chi^{\prime\prime}(\omega)$ with the Shiba's relation exactly
satisfied. Besides, the coherent-incoherent transition point
$\alpha_c$ can be determined, which has not been demonstrated for
the structured bath by previous authors up to our knowledge.

\section{Acknowledgement}
This work was supported by National Natural Science Foundation of
China (Grants No. 10474062 and No. 90503007).

\listoffigures
\end{document}